\documentclass[eqsecnum,showpacs,aps,prd,epsf,twocolumn]{revtex4}
\usepackage{amsmath,amssymb}
\usepackage{graphicx}

\newcommand{\gsim}{\mbox{\raisebox{-1.ex}{$\stackrel
      {\textstyle>}{\textstyle\sim}$}}}

\def\Vec#1{\mbox{\boldmath $#1$}}
\def\til#1{\tilde{#1}} 
\begin{document}
\thispagestyle{empty}
\title {Collision of Domain Walls and Reheating of the Brane Universe}
\author{Yu-ichi Takamizu$^{1}$}
\email{takamizu@gravity.phys.waseda.ac.jp}
\author{Kei-ichi Maeda$^{1\,,2\,,3}$}
\email{maeda@gravity.phys.waseda.ac.jp}
\address{\,\\ \,\\
$^{1}$ Department of Physics, Waseda University,
Okubo 3-4-1, Shinjuku, Tokyo 169-8555, Japan\\
$^{2}$ Advanced Research Institute for Science and Engineering,
 Waseda University,
Okubo 3-4-1, Shinjuku, Tokyo 169-8555, Japan\\
$^{3}$ Waseda Institute for Astrophysics, Waseda University,
Okubo 3-4-1, Shinjuku, Tokyo 169-8555, Japan}
\date{\today}

\begin{abstract}
We study a particle production at the collision of two domain walls in 
5-dimensional Minkowski  spacetime.
This may provide the reheating mechanism of an ekpyrotic (or cyclic) 
brane universe, in which two BPS branes collide and evolve into a hot big
bang universe. We  evaluate a production rate of particles confined to
the domain wall.
The energy density of created particles is given as 
$\rho \approx  20 \bar{g}^4 N_b ~m_\eta^4 $ where $\bar{g}$ is a coupling
constant of particles to a domain-wall scalar field, $N_b
$ is the number of bounces at  the collision and $m_\eta$ is a fundamental
mass scale of the domain wall. It does not depend on the width  $d$ of the
domain wall, although the typical energy scale of  created
particles is given by $\omega\sim 1/d$.
The  reheating temperature is evaluated as
$T_{\rm R}\approx 0.88 ~ \bar{g} ~ N_b^{1/4}$.
In order to have the baryogenesis  at the electro-weak energy scale,
the fundamental mass scale is constrained
as $m_\eta \gsim 1.1\times 10^7$ GeV for $\bar{g}\sim 10^{-5}$.

\end{abstract}

\pacs{98.80.Cq}

\maketitle
\section{\label{}introduction}
The big bang theory is very successful because 
it naturally explains the evolution of our universe from 
the nucleosynthesis to the 
present time with  many  observational data. 
However, it contains some theoretical key 
problems such as the flatness and 
the horizon problems\cite{22, 21}. So far, only the idea of inflation
 provides a resolution of those problems. 
Not only it gives some picture of 
the earlier stage of the universe before a big bang 
 but also it seems to be supported by  some of recent observational data 
on CMB.
While, it is still unclear what is the origin of inflaton.
So far, 
there is no convincing link with fundamental unified theories 
 such as a string/M  theory. 

Recently a new paradigm on the early universe has been proposed, 
which is so-called a brane world\cite{1, 3}. Such speculation has
been inspired by recent developments in
string/M-theory\cite{Witten, String, Polchinski}.  
There has been tremendous works in this scheme of
dimensional reduction where ordinary matter fields are confined to
a lower-dimensional hypersurface, while only gravitational fields
propagate throughout all of spacetime.
In particular, it was shown that  the 10-dimensional 
$E_8 \times E_8$ heterotic string theory, which is 
a strong candidate to describe our real world, 
is equivalent to an 11-dimensional M theory compactified 
to ${\bf M}^{10} \times {\bf S}^1/Z_2$ \cite{Witten}. 
Then the 10-dimensional 
spacetime is expected to be compactified into 
${\bf M}^4 \times {\bf CY}^6$, where ${\bf M}^4$ and ${\bf CY}^6$
are 4-dimensional Minkowski spacetime and 6-dimensional Calabi-Yau space,
 respectively.
Randall and Sundrum~\cite{4} also proposed a new model where
four-dimensional Newtonian gravity is recovered at low energies
even without compact extra dimensions. 
Based on such a new world picture, 
many cosmological scenario have been studied
\cite{brane_cosmology,
brane_cosmo2,Lukas}. 
See also recent reviews\cite{Maeda,Maartens,Langlois,Brax}.
We have found some 
deviations from standard cosmology by modifications of 
4-dimensional Einstein equations on the brane\cite{shiromizu}, even 
for the case there is a scalar field in bulk\cite{Maeda_Wands}. 

\ In such a  brane world scenario, for resolving 
the above-mentioned theoretical key problems
in the big bang theory,
new idea of the early universe has been proposed, which is called 
the ekpyrotic scenario or a cyclic universe scenario
\cite{Khoury,Khoury_cyclic}.
It is based on a collision of two cold branes. 
The universe starts with a cold, empty and 
nearly BPS ground state of M
theory, which contains two parallel branes at rest. 
Two branes approach each other and then collide.
The energy is 
dissipated on the brane and the big bang universe will start. 
The BPS state is required in order to remain a supersymmetry
in a low-energy 4-dimensional 
effective action. The visible and hidden 
branes are flat, which are described by a Minkowski spacetime,
 but the bulk is warped along the fifth dimension. 
Since this scenario is not only motivated by the fundamental unified 
theory but also it may resolve the theoretical key problems such as the
flatness and horizon problems, it would be very attractive. 
There are also many discussions about 
density perturbations to see 
whether this scenario is really a reliable model for the early 
universe\cite{density_ekpyrotic,density_ekpyrotic2,
density_ekpyrotic3,density_ekpyrotic4}. 

On the other hand, even though there are some works by\cite{kofman}, 
the reheating process itself in this scenario 
has not been so far  investigated in  detail.
Hence, in this paper, we study how we can recover the hot big bang universe 
after 
the collision  of the branes.
Here we investigate quantum creation of particles, which are  confined to 
the brane,  at the collision of two branes. 
It may be difficult to deal properly with  
the collision of  two branes in a  basic string theory. 
Hence, in this paper, we adopt  a domain wall which is constructed by some
scalar  field as a brane,   and analyze the collision of two domain walls in a
5-dimensional bulk spacetime.  Some works also adopt such a picture
\cite{Copeland, Dvali}. 
It is worth noting that there is a thick domain wall model for a  brane
world\cite{sakai}.

In order to analyze particle creation at the brane collision,
in this paper we consider the simplest situation. 
We discuss that
 two domain walls collide in a 5D Minkowski spacetime. 
In Sec. \ref{brane colli}, we analyze the collision of two domain walls. 
Then, in Sec. \ref{III}, we investigate particle creation on the wall 
at the collision.  Applying the particle production to the 
energy dissipation of the 
brane, we discuss the reheating mechanism of a brane universe.
We use the unit of $c=\hbar=1$.

\section{\label{brane colli}
collision of two domain walls} 
\subsection{Basic equations and initial setting}
We study the collision of two flat domain walls in
5-dimensional (5D) Minkowski spacetime.
To construct a domain wall structure, we adopt
a real scalar field
$\Phi$ 
with a double-well potential,
\begin{equation}
V(\Phi)=\frac{\lambda}{4}(\Phi^2-\eta^2)^2\,,
\label{1_}
\end{equation}
where the potential minima are located at $\Phi=\pm\eta$. 

We discuss the collision of two parallel domain walls.  The scalar field is assumed to
depend only on a time coordinate $t$ and  one spatial coordinate $y$.
 The rest three spatial
coordinates are denoted by $\Vec{x}$. 
We rescale the parameters and variables 
by $\eta$ (or its mass scale 
$m_\eta=\eta^{2/3}$) as
\begin{equation}
\ \til{t}=m_\eta t \ ,
\ \til{y}=m_\eta y \ ,
\til{\Phi}=\frac{\Phi}{\eta}\ ,
\ \til{\lambda}= m_\eta  \lambda\ .
\end{equation}
In what follows, we omit the tilde in dimensionless variables for brevity.

The equation of motion for
$\Phi$ in 5D is given by
\begin{equation}
\ddot{\Phi}
-\Phi''
+\lambda\,\Phi\,(\Phi^2-1)=0\,,
\label{8}
\end{equation}
where $\dot{}$ and $'$  denote ${\partial}/{\partial t}$ and 
 ${\partial}/{\partial y}$, respectively.

Eq. (\ref{8}) has not only two stable vacuum 
solutions $\Phi_{\pm}=\pm 1$ but also
a static kink solution (${K}$), which is topologically stable. 
The latter one is called a domain wall, which is described by 
\begin{equation}
\Phi_K(y)=\tanh\left[\ \frac{y}{d}\ \right]
\quad,\ 
\label{2_}
\end{equation}
where $d=\sqrt{{2}/{\lambda}}$ is the thickness of the wall \cite{11}. 
The antikink solution ($\bar{K}$) is obtained from Eq. (\ref{2_}) 
by reflecting the spatial coordinate $y$  as
$\Phi_{\bar{K}}(y)=\Phi_K(-y)=-\Phi_K(y)$.  
A single domain wall solution moving with
a constant speed
$\upsilon$ in the $y$  direction is given by boosting the
solution (\ref{2_}). We find a moving domain wall solution, whose initial
position 
is located at $y=0$, as
\begin{equation}
\Phi_\upsilon(y,t)=\tanh\left[\ \frac{\gamma}{d}\left(y-\upsilon
t\right)\
\right]
\ ,
\end{equation}
where $\gamma=1/\sqrt{1-\upsilon^2}$\ is the Lorentz factor.

In order to discuss the collision of two domain walls, we first have to set
up the initial data.
We put a kink and an
antikink solution, which are separated  at large distance, 
and are  moving toward each other with
velocities
$\upsilon$ and $-\upsilon$. 
The explicit initial configuration 
is given by
\begin{eqnarray}
\Phi(y,0)=
\Phi_\upsilon(y+y_0,0)-\Phi_{-\upsilon}(y-y_0,0)-1 \ ,
\label{7} 
\end{eqnarray}
where $y=\pm y_0 $ is the initial position of the  walls. 
 The  spatial
separation of two  walls is given by $2y_0$, and 
as long as the separation distance is much larger 
than the thickness of a wall ($y_0\gg d$),
the initial profile (\ref{7}) gives a good 
approximate solution for two moving domain walls. 
The initial value of $\dot{\Phi}$ is also given by
\begin{eqnarray}
\dot{\Phi}(y,0)=
\dot{\Phi}_\upsilon(y+y_0,0)-\dot{\Phi}_{-\upsilon}(y-y_0,0)\ .
\label{initialPhidot} 
\end{eqnarray}
Using this initial values, we  solve
the dynamical  equation (\ref{8}) numerically, whose results will be
shown in the next subsection.

\subsection{\label{numer} Time Evolution of Domain Walls}

\begin{figure}[ht]
\begin{center}
\includegraphics[width=8cm]{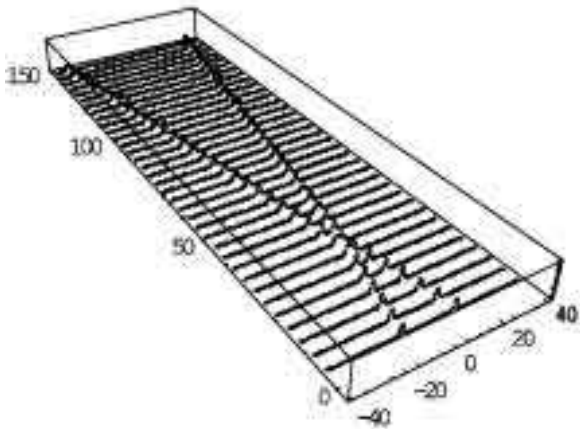}
\caption{Collision of two domain walls 
for the case of the initial velocity $\upsilon=0.4$. 
The time evolution of the energy density of scalar field 
$\Phi$ is shown from $t=0$ to 150. 
The maximum point of $\rho_{\Phi}$ defines the position
of a wall ($y=y_{\rm W}(t)$).}
\label{9} 
\end{center}  
\end{figure}

\begin{figure}[ht]
\begin{center}
\includegraphics[width=8cm]{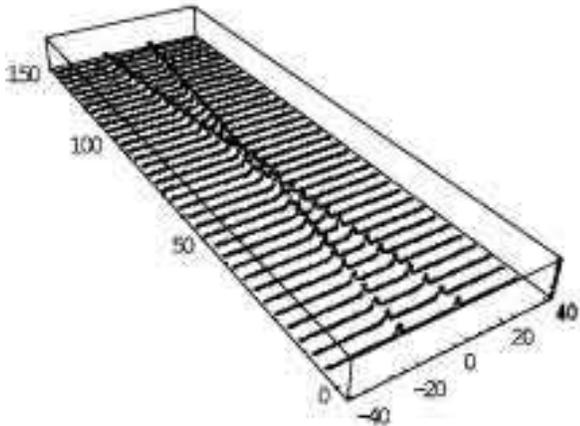}
\caption{Collision of two domain walls
for the case of the initial velocity $\upsilon=0.2$. 
The time evolution of the energy density of 
the scalar field $\Phi$ is shown from $t=0$ to 150.
We find that collision occurs twice at $t\approx 58$ and 77.
We set ${\lambda}=1.0$. 
From this figure, we find clearly that collision occurs twice.}
\label{10}	
\end{center}
\end{figure}

We use a numerical approach to solve 
the equations for the colliding domain walls.
The numerical method is shown in Appendix \ref{appen_a}. 
We have two free parameters in our
simulation of the two-wall collision, i.e.  a thickness of the wall 
${d}=\sqrt{2/{\lambda}}$ and an initial velocity of the walls
${\upsilon}$. 
The collision of two walls
was discussed by \cite{fractal} in 4-dimensional Minkowski space.
Although we discuss the domain wall collision in 5-dimensional Minkowski
space, our basic equations are exactly the same as their case, and then
we find the same results as theirs.
In particular, the results are very sensitive to  the initial
velocity
$\upsilon$. 

\ First we show the numerical results for two typical initial velocities, i.e.
$\upsilon =0.2$ and 0.4 in Figs. \ref{9}, \ref{10}. The 
evolution of the energy density of $\Phi$ is shown in Figs. \ref{9} and
\ref{10}.  The energy density of the scalar field
is given by
\begin{equation}
\rho_{\Phi}=\frac{1}{2}
\left[\dot{\Phi}^2+{\Phi'}^2
+\frac{\lambda}{2}(\Phi^2-1)^2\right]\,.
\label{11_}
\end{equation}
From Figs. \ref{9} and \ref{10}, 
we find some peaks of the energy density, by which  we
can define the positions of moving walls ($y=\pm y_{\rm W}(t)$).

\ From Figs. \ref{9} and \ref{10}, 
we see the detail of the collision as follows.
For the case of initial velocity $\upsilon=0.4$,
we find that, the collision occurs once, while 
it does twice for the case of  $\upsilon=0.2$.
To be precise, in the latter case, after two  walls 
collide first, they bounce, recede to a finite distance, and then return
to collide again. \\

As shown by several authors \cite{fractal,13,14,15}, however, 
the result highly depends on the
incident velocity  $\upsilon$  .
In Appendix B, we show our detail analysis, which confirms the previous 
works.
For a sufficiently large velocity, it is expected that 
a kink and antikink will just bounce off once because 
there is no time to exchange the energy during the collisional 
process. 
In fact, it was shown in \cite{fractal} that
two walls just bounce off once 
for $\upsilon \gsim 0.25$. 
For smaller velocity, we find multiple bounces when they collide.
The number of bounce
during the collision depends complicatedly on  the incident velocity.
For example, the bounce occurs once for  $\upsilon=0.4$, while  twice
for
$\upsilon=0.2$.
We also find many bounce solutions for other incident velocities as
shown in Appendix B (see also \cite{fractal}).
We find that the number of bounce is not monotonic function of $\upsilon$,
but is much more complicated.
If we change the incident velocity slightly, the number of bounce 
changes. The existence of a fractal structure in the parameter space is
shown in Fig. 6 of
\cite{fractal}.

\section{\label{III}
Particle production on a moving domain wall}
\subsection{\label{reheat}
Quantization of a particle on the domain wall and its
production rate}

Once we find the solution of colliding domain walls, we can evaluate
the time evolution of a scalar field on the domain wall. 
Since we assume that we are living on one domain wall, we are interested 
in  production of a particle confined to the  domain wall.
We assume that there is some coupling between a 5D scalar field 
$\Phi$ which is responsible for the domain wall and a particle on the 
domain wall.
Because the value of the scalar field changes with time, we expect
quantum particle production occurs. 
This may be important for a reheating mechanism of the
colliding domain walls.

Hence we have to know the value of the 
scalar filed $\Phi$ on the domain wall, i.e.
$\Phi_{\rm W}(\tau)=\Phi(t, y_{\rm W}(t))$.
Since the wall is moving in a 5D Minkowski space, 
we have to use the proper time $\tau$ of the wall,
which is given by 
\begin{equation}
\tau=\int^t_0  dt \sqrt{1-\dot{y}_{\rm W}^{~2}(t)} \,\,,
\end{equation}
 when we estimate the
particle production in our 4-dimensional  domain wall.

\begin{figure}[ht]
 \centering
 \includegraphics[width=8cm]{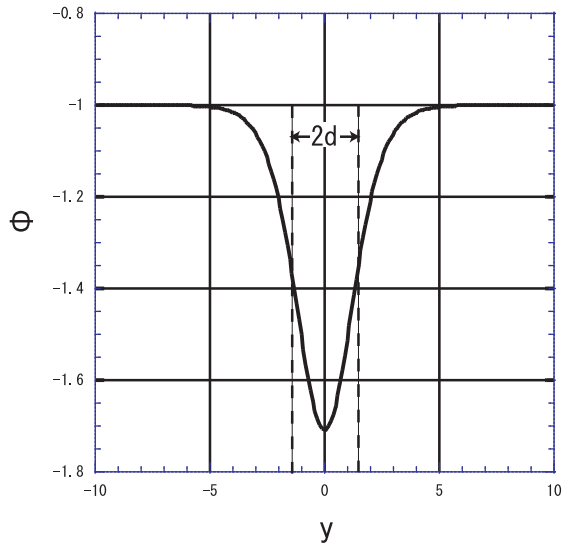}
\caption[fig1]{The spatial distribution of the scalar  field $\Phi$ when
the domain walls collide. $d$ is the thickness of the wall.
The scalar field has non-vanishing value for the effective width  of $2d$
at the collision. }
\label{thickness}
\end{figure}

We  consider a particle on the domain wall 
which is described by a scalar field  
$\psi$. We assume that it  is confined  to the domain wall and couples
to  the scalar field
$\Phi$  as
\begin{eqnarray}
g^2\,\phi^2(\tau) \psi^2\,, 
\end{eqnarray}
where $g$ is a coupling constant.
Here $\phi(\tau)$ is the 4-dimensional scalar field, which is 
induced from the 5-dimensional $\Phi$ as 
\begin{equation}
\phi(\tau) \approx \sqrt{2d} ~
\Phi_{\rm W}(\tau)\ .
\end{equation}
We have also assumed that the effective width of the domain 
wall  when it
collides is about
$2d$, which is confirmed by numerical simulation.
In Fig. \ref{thickness}, we depict the spatial  distribution of
the scalar field
$\Phi$ when the domain walls collide.
Note that $\Phi_{\rm w}(\tau)$ is the maximum amplitude  in this
distribution.

The basic equation for
$\psi$ is then  
\begin{equation}
\frac{\partial^2\psi}{\partial \tau^2}
-\nabla^2\psi
+\bar{g}^2\,\Phi_{\rm W}^2(\tau)\,\psi=0\label{1}\ ,
\end{equation}
where 
\begin{equation}
\bar{g}^2=2 d g^2 ={2\sqrt{2}\over \lambda^{1/2}} \, g^2 \ .
\end{equation}
Since our background spacetime is the 4-dimensional Minkowski
space, it is easy to quantize the scalar field $\psi$.
In a canonical quantization scheme
\cite{quantum}, 
we expand $\psi$ as  
\begin{equation}
\psi(\tau,\Vec{x})=\sum_{k}[a_k\,\psi_k(\tau)\,u_k(\Vec{x})
 +a_k^{\dag}\psi_k^{*}(\tau)u_k^{*}(\Vec{x})]
\label{2}\ ,
\end{equation}
where\ $u_k(\Vec{x})=(2\pi)^{-\frac{3}{2}} e^{i\Vec{k}\cdot\Vec{x}}$.
The wave equation (\ref{1}) for each mode is now
\begin{equation}
\ddot{\psi_k}
+[k^2+\bar{g}^2\Phi_{\rm W}^2(\tau)]\psi_k
=0\ ,
\label{3_}
\end{equation}
where $k=|\Vec{k}|$.
Since two domain walls are initially far away each other,
the value of $\Phi_{\rm W} $ is almost zero.
We can quantize $\psi$ as a usual  quantization scheme.
The eigen function with a positive frequency
is given by
\begin{equation}
\psi_{k}^{({\rm in})}={1\over \sqrt{2\omega_k}}e^{-i\omega_k \tau}\ ,
\label{4_}
\end{equation}
where\ $\omega_k=\sqrt{k^2+\bar{g}^2\Phi_{\rm W}^2(0)}\approx k$. 
We impose the equal time commutation relation for the operators 
$a_k$ and $a_k^{\dag}$ 
\begin{eqnarray}
&&[a_k, a_{k'}]= 0\ ,\\
&&[a_k ^{\dag},a_{k'} ^{\dag}] = 0\ ,\\
&&[a_k,a_{k'}^{\dag}] =\delta_{k k'}
\ ,
\end{eqnarray}
where $a_k$ and $a_k^{\dag}$ are an annihilation and 
a creation operators.
We then define a vacuum $|0>_{\rm in}$ at $\tau=0$ by $a_k$ as
\begin{equation}
a_k|0>_{\rm in}=0\quad\ ,\ \forall k\ .
\end{equation} 
After the collision of domain walls, we expect that the value of  $\Phi_{\rm
W}$ again approaches zero (see the detail in the next subsection). We can
also define the vacuum state $|0>_{\rm out}$, which is different from the initial
vacuum state $|0>_{\rm in}$. The eigen function of $\psi_k$ for $\tau\rightarrow \infty$
is then given by a linear combination of $\psi_{k}^{({\rm in})}$ and
$\psi_{k}^{({\rm in})*}$ as 
\begin{eqnarray}
\psi_{k}^{({\rm out})}=\alpha_k\,\psi_{k}^{({\rm in})}
+\beta_k\,\psi_{k}^{({\rm in})*}\ , 
\label{6_}
\end{eqnarray}
and the annihilation and creation operators as
\begin{equation}
\bar{a}_k=\alpha_k a_k+\beta_k^{*} a_k^{\dag}\ ,
\label{5}
\end{equation}
where\ $\alpha_k,\ \beta_k$\ are the Bogolubov coefficients,
 which satisfy
the normalization condition
\begin{equation}
|\alpha_k|^2-|\beta_k|^2=1\ .
\end{equation}
The Hamiltonian of this system is given by
\begin{eqnarray}
:H:&=&-\int_{\tau={\rm const}} :T^{0}_{~~0}:\,d^3\Vec{x}\nonumber \\
&=&\sum_k a_k^{\dag} a_k\omega_k\ ,
\label{4}
\end{eqnarray}
where :\ : is the normal ordering operation.
The creation of the particles with mode $k$ is evaluated as
\begin{equation}
<0|_{\rm in}:H_k:|0>_{\rm in}=|\beta_k|^2\omega_k \ {\rm as}\  \tau \rightarrow
\infty\ .
\end{equation}
As a result, the number density and energy density 
of produced particles are given by 
\begin{eqnarray}
n&=&\int |\beta_k|^2d^3\Vec{k}\ ,
\label{13_}\\
\rho&=&\int |\beta_k|^2 \omega_k
d^3\Vec{k}\ \label{14_}.
\end{eqnarray}

\subsection{Time evolution  of a scalar field on the domain wall 
and particle production\label{reheat2}}
\begin{figure}[ht]
 \centering
 \includegraphics[width=8cm]{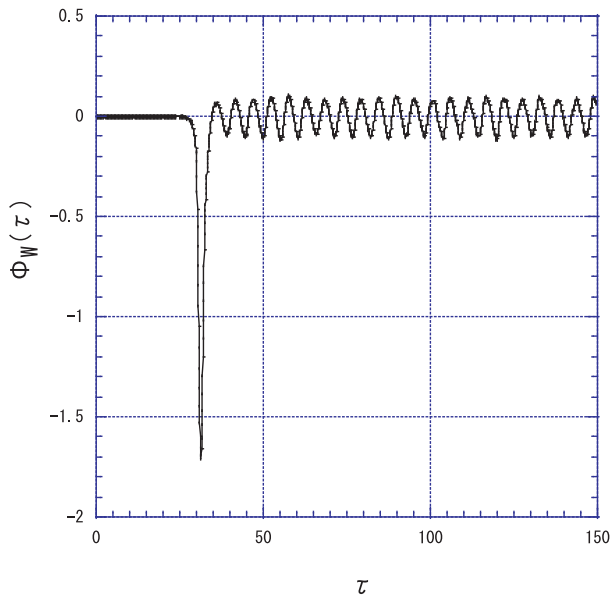}
\caption[fig1]{Time evolution of a scalar field 
on one moving wall 
for $\upsilon=0.4,\ {\lambda}=1.0$.
The value of the scalar field is given by  $\Phi_{\rm W}(\tau)
=\Phi(t, y_{\rm W}(t))$, where $y_{\rm W}(t)$ is the position of the wall
and $\tau$ is the proper time on the wall.
} \label{11}
\end{figure}

\begin{figure}[ht]
 \centering
 \includegraphics[width=8cm]{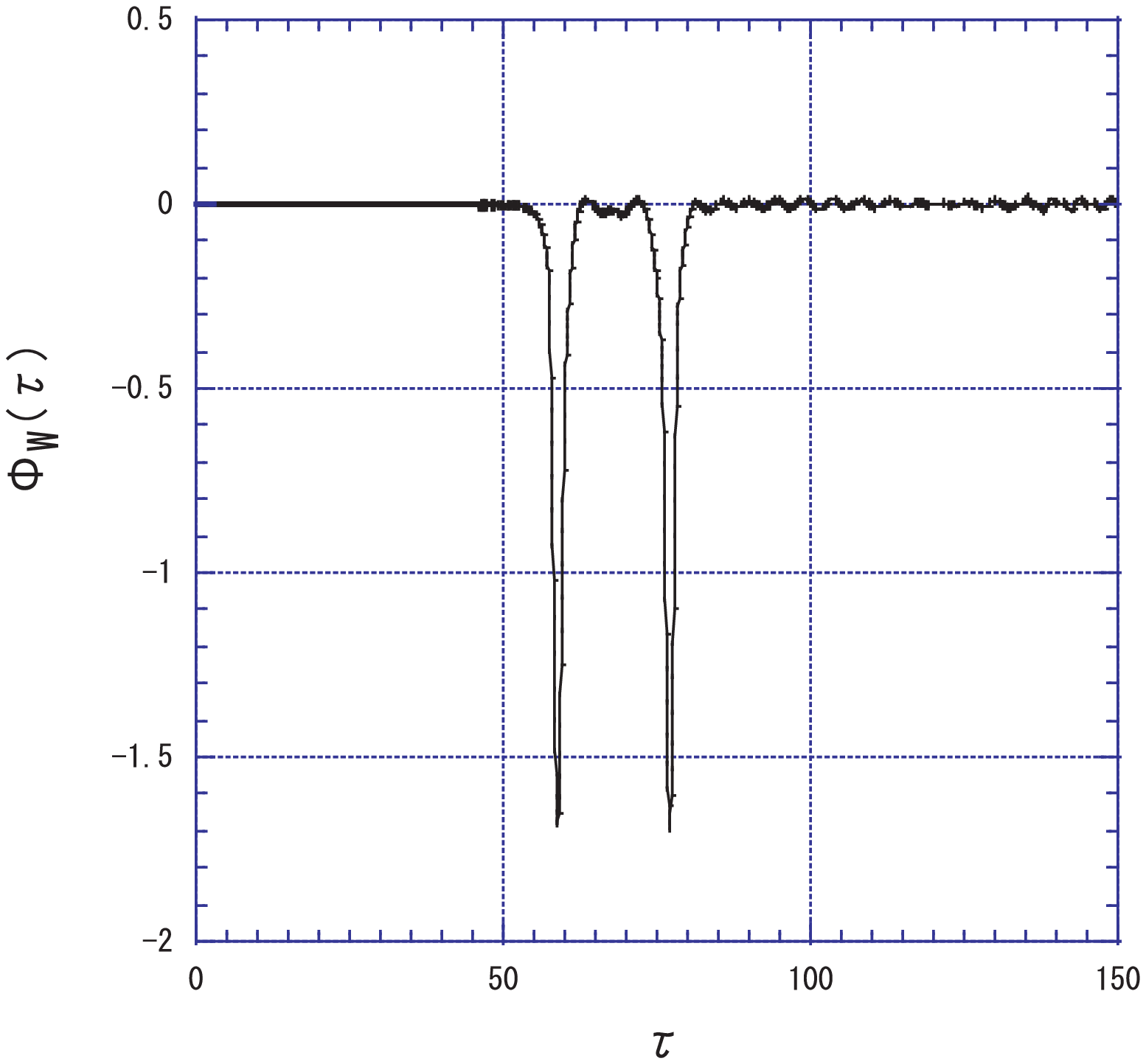}
\caption[fig1]{
Time evolution of a scalar field $\Phi_{\rm W}(\tau)$ on one moving wall 
for $\upsilon=0.2,\ {\lambda}=1.0$.
} \label{12}
\end{figure}

Now we estimate the particle production by the domain wall collision.
In Figs. \ref{11} and \ref{12}, we depict the time
evolution of
$\Phi_{\rm W}$ on one moving wall  with respect to $\tau$. 
In Fig. \ref{9}, 
we find one collision point, 
which corresponds to  a spike in Fig. \ref{11},
and two-bounce in  Fig. \ref{10}
gives two spikes in Fig. \ref{12}.

\begin{figure}[hb]
 \centering
 \includegraphics[width=8cm]{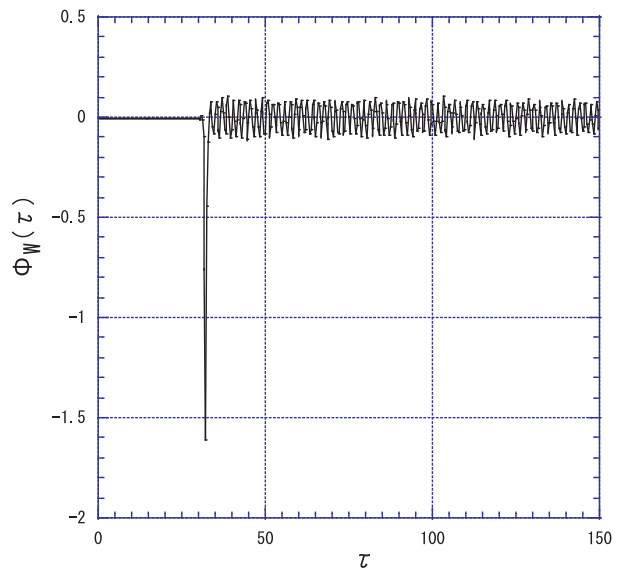}
\caption[fig1]{
Time evolution of a scalar field $\Phi_{\rm W}(\tau)$ on one moving wall 
for $\upsilon=0.4,\ {\lambda}=10$.
} \label{11.5}
\end{figure}

\begin{figure}[ht]
 \centering
 \includegraphics[width=8cm]{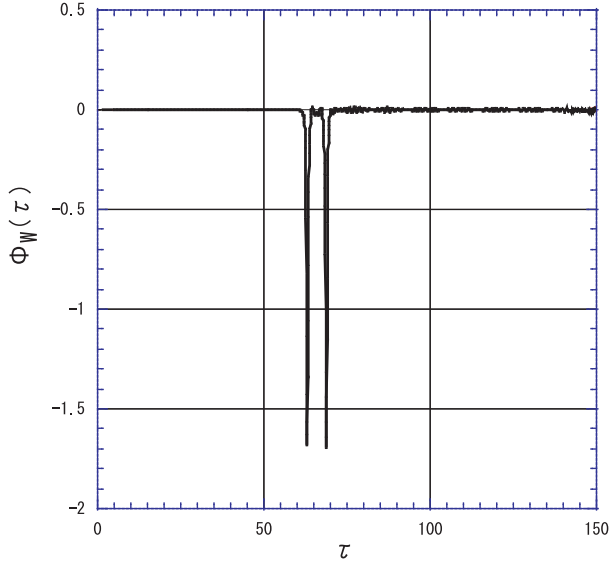}
\caption[fig1]{
Time evolution of a scalar field $\Phi_{\rm W}(\tau)$ on one moving wall 
for $\upsilon=0.2,\ {\lambda}=10$.
} \label{13}
\end{figure}

We also show the results for different values of  the coupling
constant 
$\lambda$ in Figs. \ref{11.5} and \ref{13}
($\lambda=10$). If Fig. \ref{11.5}  we find that
when $\lambda$ gets larger as
$\lambda=10$,  the spike of $\Phi_{\rm W}$ becomes  sharp. 
The same thing happens in the case of two bounce (see Fig. 
\ref{13}). 

In Figs. \ref{11}-\ref{13}, we find  
that $\Phi_{\rm W}$ begins to oscillate after the collision. 
We also find that 
the period of these oscillations in 
Figs. \ref{11.5} and \ref{13} is shorter that those in Figs.  \ref{11} and
\ref{12}. 
One may wonder whether this oscillation is realistic or not.
This oscillation, however,  turns out
not to be a numerical error but a real oscillation of the domain wall.
In
Appendix \ref{perturb},  using  perturbation analysis
we show there is one stable oscillation around the kink solution $\Phi_K(y)$.
We expect that the oscillation is excited by the collision.
In fact, the amplitude of the oscillation increases as the incident velocity 
$\upsilon$ increases.
In large velocity limit ($\upsilon \gsim 0.6$),
 we find $\Phi_\infty^2\approx
0.18(\gamma-1)$, where  $\Phi_\infty$ is the amplitude of the post-oscillation.

Since the scalar field on the domain wall oscillates
as $\Phi_{\rm W}\approx \Phi_\infty \cos \sigma \tau$ after the collision,
our wave equation (\ref{3_})  would be rewritten as
\begin{equation}
\ddot{\psi_k}
+\left[k^2+{1\over 2} 
\bar{g}^2\Phi_\infty^2\left(1+\cos 2\sigma \tau
\right)
\right]\psi_k
=0\ ,\label{3_1}
\end{equation}
where $\sigma=\sqrt{3/2}\lambda^{1/2}$ is the eigenvalue of  the
perturbation eigen function, it is so-called Methieu equation. 

From this equation, we may wonder whether we can ignore this  oscillation
when we evaluate particle production rate.
In fact, when we discuss a preheating mechanism via a parametric
resonance with a similar oscillating behavior\cite{preheating}, 
in conventional cosmology\cite{22} as well as brane world cosmology
\cite{tujikawa}. 
It is known that for Methieu equation, 
there is an exponential instability ${\psi_k}\propto \exp(\mu_k t)$
within a set of resonance bands, where $\mu_k=\bar{g}\Phi^2_{\infty}/8$. 
 This instability corresponds to  an
exponential growth of created particles, which is essential 
in the preheating mechanism. In order to get a successful particle
production by this resonance instability, however, we have to require large value
of $\bar{g}^2\Phi_\infty^2$. However, in the present simulation, it is
rather small, e.g.
$\Phi_\infty\sim 0.1$ for the case of $\upsilon =0.4$.
Hence, we may ignore such a particle production by a parametric  resonance
in the present calculation.  However, if the incident velocity is very 
fast such as a speed of light, we may find a large oscillation. Then we
could have an instant preheating process by the domain wall collision.
We also wonder whether or not the standard reheating mechanism due to the decay of
oscillating scalar field is effective.
In this case, we have to evaluate the decay rate $\Gamma_\phi$ to other particles.
Since $\Gamma_\phi \propto \bar{g}^4$, we expect that the reheating temperature is
proportional to $\bar{g}^2$, which is enough small to be ignored.
Note that there is another factor which reduces the decay rate in the case that
the potential is not a spontaneous symmetry breaking type\cite{preheating}.

 In what follows, we ignore the creation due to post-oscillation stage.
Hence, we just follow the procedure shown in the previous subsection.
 Using the evolution of the scalar field $\Phi_{\rm W}$, we calculate the 
Bogolubov  coefficients $\alpha_k$ and $\beta_k$.
In Table \ref{6}, we show the results for three different parameters; 
$\upsilon$ (the incident velocity), $\lambda$ (the self-coupling constant 
of the scalar field), and $\bar{g}$
(the coupling constant to a particle $\psi$).

\begin{table}[t]
	\begin{center}
		\begin{tabular}{|c|c|c|c|c|c|c|}
		\hline
 {$\bar{g}$} &{$\upsilon$} & {${\lambda}$}
	&$d$	& {$N_b$} 
		& {$n$} & {$\rho$}  \\
		\hline
& \raisebox{-1.5ex}[0pt]{0.4} & 1.0 &1.414& \raisebox{-1.5ex}[0pt]{1}
		& 3.69$\times 10^{-7}$
		& 2.05$\times 10^{-7}$
	\\
\cline{3-4} 
\cline{6-7} 
\raisebox{-1.5ex}[0pt]{$0.01$}	 && 10 & 0.447&
		& 1.16$\times 10^{-7}$
		& 2.05$\times 10^{-7}$
\\
\cline{2-7} 
	& \raisebox{-1.5ex}[0pt]{0.2}  & 1.0 & 1.414 & \raisebox{-1.5ex}[0pt]{2} 
		& 7.19$\times 10^{-7}$
		& 3.90$\times 10^{-7}$
\\
\cline{3-4} 
\cline{6-7} 
	& & 10 &0.447   &
		& 2.26$\times 10^{-7}$
		& 3.91$\times 10^{-7}$
\\
		\hline
& \raisebox{-1.5ex}[0pt]{0.4} & 1.0 &1.414  &\raisebox{-1.5ex}[0pt]{1}  
		& 3.57$\times 10^{-3}$
		& 2.01$\times 10^{-3}$
\\
\cline{3-4} 
\cline{6-7} 
\raisebox{-1.5ex}[0pt]{0.1}   &   & 10  &0.447 &
		& 1.16$\times 10^{-3}$
		& 2.05$\times 10^{-3}$
\\
	\cline{2-7} 
& \raisebox{-1.5ex}[0pt]{0.2} & 1.0 &1.414  & \raisebox{-1.5ex}[0pt]{2} 
		& 6.65$\times 10^{-3}$
		& 3.81$\times 10^{-3}$
\\
\cline{3-4} 
\cline{6-7} 
 & &  10 & 0.447 &
		& 2.24$\times 10^{-3}$
		& 3.88$\times 10^{-3}$
\\
		\hline
		\end{tabular}
	\end{center}
	\caption{The number and energy densities ($n$ and $\rho$)
of created particles for the 
typical values of the coupling
$\bar{g}$, the incident velocity $\upsilon$ and the self-coupling
$\lambda$. $d=\sqrt{2/\lambda}$ and $N_b$ denote the width of the wall and
 the number of bounces 
at the collision, respectively.}
	\label{6}
\end{table}

From this Table, we find the following three features:\\
(1) The produced energy density $\rho$ depends very much on $\bar{g}$.
We study two cases with 
$\bar{g}=0.01$ and $0.1$.
The energy density for  $\bar{g}=0.1$ is $10^4$ times larger than that 
for 
$\bar{g}=0.1$, which means that $\rho$ is proportional to $\bar{g}^4$.
\\ 
(2) The energy density $\rho$ for $\upsilon=0.2$ is twice larger than that 
 for $\upsilon=0.4$. It may be so because the bounce occurs twice for 
$\upsilon=0.2$, while once for  $\upsilon=0.4$.
\\
(3) The energy density is less sensitive to $\lambda$.

We also investigate  for several different 
initial velocities because the collisional process is very sensitive to 
its incident velocity. 
We analyze many cases with
two bounces,  with three bounces, with four bounces, $\cdots$
in the range $\upsilon=0.2$-$0.25$ as shown in Appendix B. 
Using those numerical data, we also evaluate the number and energy 
densities of  the particles created at the collision.
The results are summarized  in
Table \ref{23}.
 
From this Table, we confirm the above three features (1)-(3).
In particular, it becomes more clear that
the energy density is proportional to the
number of bounce
$N_b$.

We can summarize our results by
the following empirical formula
\begin{eqnarray}
n &\approx &  25 d \bar{g}^4 N_b\ ,
\label{n_density}\\
\rho &\approx & 20 \bar{g}^4 N_b 
\label{e_density}
\ .
\end{eqnarray}

If this energy of the particles is thermalized by
interaction and thermal equilibrium state is realized,
we can estimate the reheating temperature
by
\begin{eqnarray}
\rho={\pi^2\over 30} g_{\rm eff} T_{\rm R}^4 \, ,
\label{21}
\end{eqnarray}
where
$g_{\rm eff}$ is
the effective number of  degrees of freedom of particles. 
Hence we find the reheating temperature by the domain wall collision 
as
\begin{eqnarray}
T_{\rm R}&=&\left({\pi^2\over 30}\right)^{-1/4} g_{\rm eff}^{-1/4} 
\rho^{1/4}\nonumber\\ &\approx& 0.88 \times  \left({g_{\rm eff}\over
100}\right)^{-1/4} ~\bar{g} ~ N_b^{1/4}
\,.\label{28}
\end{eqnarray}

\begin{table}[b]
	\begin{center}
		\begin{tabular}{|c|c|c|c|c|c|}
		\hline
$N_b$& {$\upsilon$} & {${\lambda}$}&$d$
	 &  {$n$} 
		& {$\rho$}  \\
		\hline
	 & \raisebox{-1.5ex}[0pt]{0.225} & 1.0  & 1.414
		& 7.03$\times 10^{-7}$
		& 3.72$\times 10^{-7}$
		\\
		\cline{3-6}
		2 & & 10& 0.447   
		& 2.21$\times 10^{-7}$
		& 3.71$\times 10^{-7}$
		\\
		\cline{2-6}
	 & \raisebox{-1.5ex}[0pt]{0.238} & 1.0  & 1.414 
		& 7.08$\times 10^{-7}$
		& 3.78$\times 10^{-7}$
		\\
		\cline{3-6}
		 &  & 10 & 0.447  
		& 2.23$\times 10^{-7}$
		& 3.78$\times 10^{-7}$
\\
		\hline
	  & \raisebox{-1.5ex}[0pt]{0.2062} & 1.0  & 1.414 
		& 1.10$\times 10^{-6}$
		& 6.07$\times 10^{-7}$
		\\
		\cline{3-6}
	 &  & 10  & 0.447
		& 3.45$\times 10^{-7}$
		& 6.06$\times 10^{-7}$
\\
		\cline{2-6}
 & \raisebox{-1.5ex}[0pt]{0.2049} & 1.0  & 1.414  
		& 1.09$\times 10^{-6}$
		& 6.01$\times 10^{-7}$
	\\
		\cline{3-6}
	3 &  & 10  & 0.447   
		& 3.43$\times 10^{-7}$
		& 6.01$\times 10^{-7}$
	\\
		\cline{2-6}
		 & \raisebox{-1.5ex}[0pt]{0.2298} & 1.0  & 1.414 
		& 1.10$\times 10^{-6}$
		& 6.04$\times 10^{-7}$
\\
		\cline{3-6}
	 &  & 10  & 0.447
		& 3.43$\times 10^{-7}$
		& 6.02$\times 10^{-7}$
	\\
		\cline{2-6}
	 & \raisebox{-1.5ex}[0pt]{0.22933} & 1.0  & 1.414
		& 1.09$\times 10^{-6}$
		& 6.03$\times 10^{-7}$
\\
		\cline{3-6}
	 &  & 10    & 0.447
		& 3.44$\times 10^{-7}$
		& 6.01$\times 10^{-7}$
\\
		\hline
	 & \raisebox{-1.5ex}[0pt]{0.229283} & 1.0  & 1.414 
		& 1.47$\times 10^{-6}$
		& 8.10$\times 10^{-7}$
	\\
		\cline{3-6}
		4 &  & 10  & 0.447
		& 4.61$\times 10^{-7}$
		& 8.09$\times 10^{-7}$
	\\
		\cline{2-6}
		 & \raisebox{-1.5ex}[0pt]{0.2292928} & 1.0  & 1.414
		& 1.47$\times 10^{-6}$
		& 8.16$\times 10^{-7}$
\\
		\cline{3-6}
	 &  & 10  & 0.447  
		& 4.62$\times 10^{-7}$
		& 8.17$\times 10^{-7}$
\\
		\hline
		\end{tabular}
	\end{center}
	\caption{The number and energy densities ($n$ and $\rho$)
 of created  particles with
respect to the number of bounces  $N_b$.
$\upsilon$ and $d$ are the incident velocity and the width of  the wall,
respectively. We set $\bar{g}=0.01$.}
	\label{23}
\end{table}

\begin{figure}[ht]
 \centering
 \includegraphics[width=8cm]{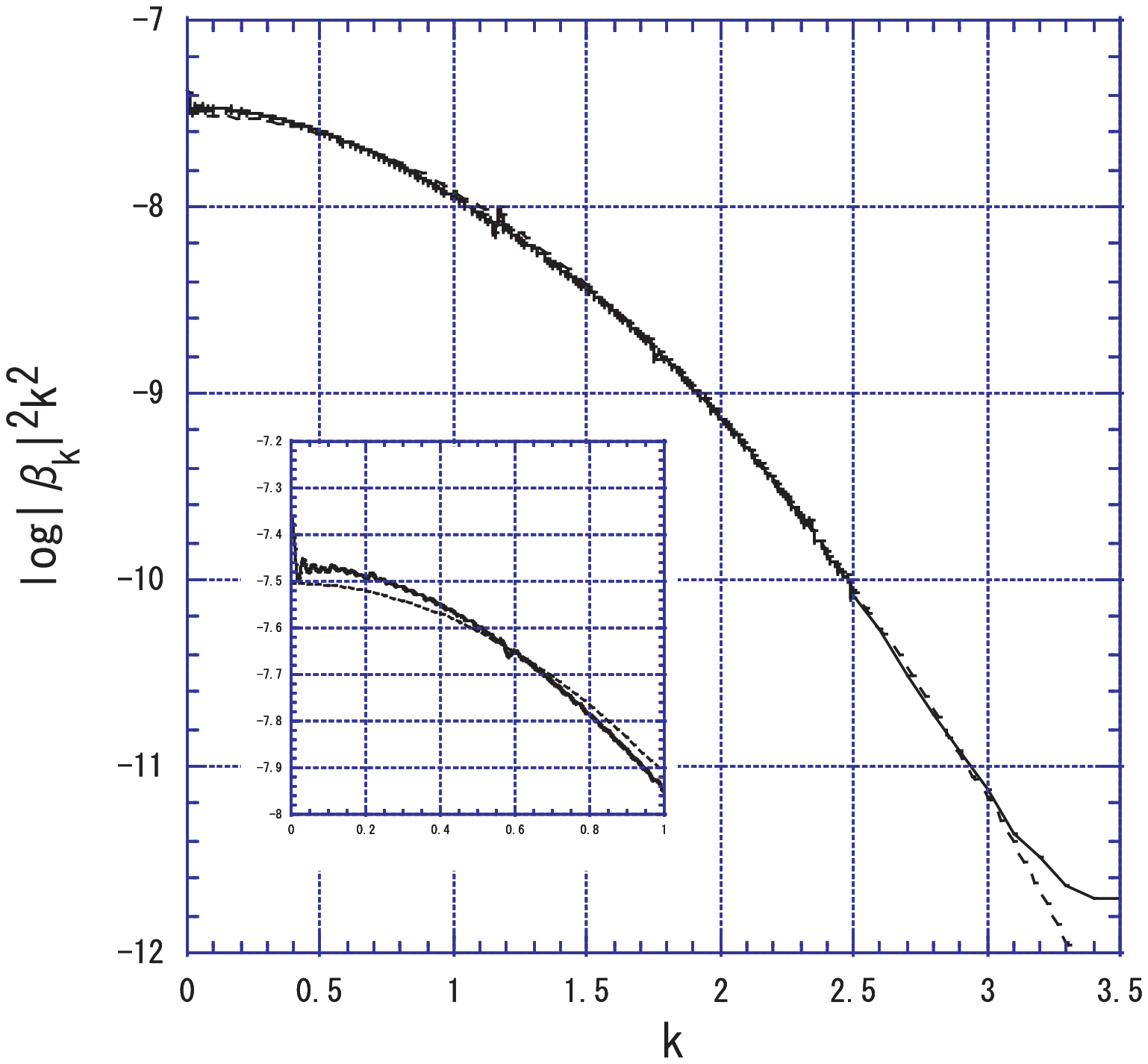}
\caption[fig1]{Spectrum of the created particles at the collision.
We plot  
$\log|\beta_k|^2k^2$
with respect to $k$
for $\upsilon=0.4,\ {\lambda}=1.0, 
{\bar{g}}=10^{-2}$.
The gaussian distribution is plotted by a dotted line,
which gives a good approximation for $k \leq 3$.
In the small box, we
enlarge the low frequency region ($k\leq 1$) to see the deviation from the
gaussian distribution.} \label{16}
\end{figure}
\begin{figure}[ht]
 \centering
 \includegraphics[width=8cm]{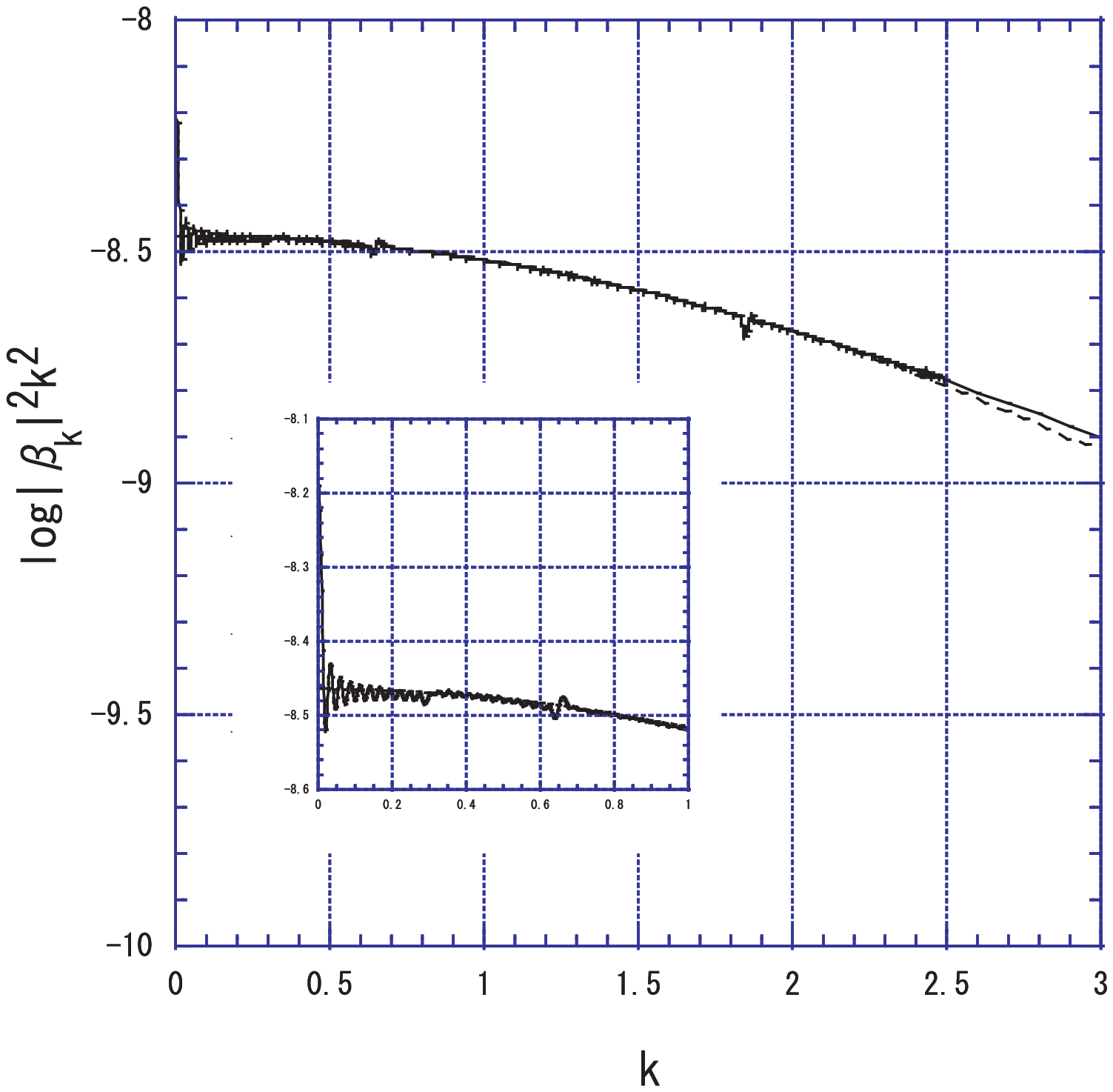}
\caption[fig1]{Spectrum of the created particles at the collision.
We plot  
$\log|\beta_k|^2k^2$
with respect to $k$
for $\upsilon=0.4,\ {\lambda}=10, 
{\bar{g}}=10^{-2}$. 
The gaussian distribution is plotted by a dotted line,
which gives a good approximation for $k \leq 2.5$.
In the small box, we
enlarge the low frequency region ($k\leq 1$) to see the deviation from the
gaussian distribution.}
\label{18}
\end{figure}

In order to see more details, in Figs. \ref{16} and \ref{18}, 
 we show a spectrum
of the produced particles of  number density $n$, i.e.
\begin{eqnarray}
n=\int_0^\infty  dk ~n_k
~~{\rm with} ~~n_k=4\pi |\beta_k|^2k^2\ .
\end{eqnarray}
The spectrum $n_k$ is well fitted as a gaussian distribution as
\begin{equation}
n_k\approx 4\pi A e^{-{k^2\over 2 k_0^2}}
\label{spectrum}
\,,
\end{equation}
where $k_0=0.73$ and $A=3.12 \times 10^{-8}$ for Fig. 10 and 
 $k_0=2.04$ and $A=3.43 \times 10^{-9}$ for Fig. 11, although there is 
small deviation partially.
These parameters can be described by physical quantities
as  $k_0\approx 1/d$ and $A\approx \Phi_0 d^2 \bar{g}^4$.
The reason is well understood.
 $k_0\approx 1/d$ means that the typical wave number is
given by the width of the scalar field when domain walls collide
(see Fig. \ref{thickness}).  As for $\beta$, it corresponds to the "reflection"
 coefficient of the 
"potential" given by Figs. \ref{11}-\ref{13} and then it will be 
proportional to the coupling constant $\bar{g}^2$,
and
 the reflection rate ($|\beta|^2$)
will be related to the potential depth $\Phi_0$ and the square of the
width $d^2$.

Integrating the fitting spectrum (\ref{spectrum}), we obtain 
\begin{eqnarray}
n&=&\int_0^\infty n_k dk =(2\pi)^{3/2}\Phi_0 d \bar{g}^4
\approx 25 d \bar{g}^4\ ,
\\
\rho&=& \int_0^\infty n_k \omega_k dk =
4\pi \Phi_0 \bar{g}^4 \approx 20  \bar{g}^4  
\ ,
\end{eqnarray}
which values are exactly the same as those obtained by  numerical
integration in 
 the case with one bounce (see Eqs. (\ref{n_density}) and
(\ref{e_density})). We expect that they are enhanced by the factor $N_b$
when we  find $N_b$ bounces at the  collision.

Therefore, although we obtain the particle creation numerically,
the result is easily understood and summarized by a simple formula.

\section{summary and discussion}

We study the particle production at the collision of two domain walls in  5D
Minkowski  spacetime.
This may provide the reheating mechanism of an ekpyrotic (or cyclic) 
brane universe, in which two BPS branes collide and evolve into a hot big
bang universe. We  evaluate a production rate of particles confined to
the domain wall. The energy density of created particles is approximate as
$\rho\approx  4\pi \Phi_0 \bar{g}^4 N_b $ where $\Phi_0$ is the
maximum amplitude of $\Phi_{\rm W}$, $N_b
$ is the number of bounce at the collision and  $\bar{g}$ is a coupling
constant of a particle to the scalar field of a domain wall. If this
energy is converted into standard matter fields, we find the  reheating
temperature as
$T_{\rm R}\approx 0.88 \times \bar{g} ~ N_b^{1/4}\left({g_{\rm eff}/
100}\right)^{-1/4} $.
We find that the particle creation is affected more greatly 
by a coupling constant $\bar{g}$ than the other 
two parameters $\upsilon$ and $\lambda$. 
The initial velocity 
changes the collision process, that is 
the number of bounce at the collision, but this
is less sensitive to the temperature. 
The thickness of a domain wall $d$ (or a self-coupling constant 
$\lambda$) changes the width of 
potential $\Phi_{\rm w}^2(\tau)$ 
of a particle field ($\psi$), and it changes a typical energy scale of 
created particles, which is estimated as $\omega\sim 1/d$.

In order to find a successful reheating, a reheating temperature 
must be higher than $10^2$ GeV, because we wish to explain 
the baryon number generation at the electro-weak energy scale\cite{21}.
Since Eq. (\ref{28})
is written in the following form;
\begin{align}
m_\eta &\approx 1.1 N_b^{-1/4} \bar{g}^{-1}
{T_{\rm R} }\nonumber \\
&\approx 1.1\times 10^7  [{\rm GeV}]\  N_b^{-1/4}\left({\bar{g}\over 
10^{-5}}\right)^{-1}
\left({T_{\rm R}\over
10^2 {\rm GeV}}\right)\ ,
\end{align}
we find
a constraint on a fundamental energy scale 
$m_\eta$ as
$m_\eta\gsim  1.1\times 10^7 ~{\rm GeV}$ for $\bar{g}=10^{-5}$ and
 $m_\eta\gsim 1.1\times 10^4 ~{\rm GeV}$ for $\bar{g}=10^{-2}$,
which are slightly larger than TeV scale.
Here we assume ${g_{\rm eff}}=100$

In the present work, we consider the 3-dimensional domain walls in  the
5-dimensional Minkowski space
and show that 
the particle production at the 
two-wall collision may provide a successful  mechanism for reheating in 
the ekpyrotic universe. 
In a string/M theory, however, we expect higher dimensions, e.g. 10 or 11.
If we compactify it to the effective 5-dimensional spacetime,  our work
can be applicable to such a mode.
Moreover, if we discuss a collision of a
$p$-dimensional walls (branes) in
$(p+2)$-dimensional spacetime, 
our approach can also be extended. 

In this paper, we have not taken account of effects from 
a background spacetime. 
We are planning to study how such a generalization affects
the present results about particle creation at the collision. 


\acknowledgments

We would like to thank S. Mizuno, T. Torii and D. Wands 
for useful discussions.
This work was partially supported by the Grant-in-Aid for
Scientific Research  Fund of the MEXT
 (No. 14540281) and by the Waseda University Grant for
Special Research Projects and  for The 21st Century 
COE Program (Holistic Research and Education Center for Physics 
Self-organization Systems) at Waseda University.
\appendix
\section{\label{appen_a}Numerical method}
For our numerical analysis of the domain wall collision, 
we solve the partial differential equation (\ref{8}) 
on discrete spatial grids with
a periodic  boundary condition. 
The scalar field on the grid points is defined by 
${\Phi}_n(t)={\Phi}({y}_n,t)$, where ${y}_n=n\triangle y$, for 
$n=1,2,\ldots ,N$. 
We use the fourth-order center difference scheme to 
approximate the second spatial derivative \cite{12} as 
\begin{align}
\frac{\partial^2{\Phi}_n}{\partial{y}^2}=
&\frac{1}{12(\triangle {y})^2}
\left[-{\Phi}_{n-2}+16{\Phi}_{n-1}-30{\Phi}_n
\right.
\nonumber\\
&\left.
~~~~~+16{\Phi}_{n+1}-{\Phi}_{n+2}\right]+O((\triangle {y})^4)\ .
\end{align}
This leads to a set of $N$ coupled second-order ordinary differential 
equation (ODE's) for the ${\Phi}_n$, i.e. 
\begin{align}
\frac{d^2{\Phi}_n}{d{t}^2}=
&&\frac{1}{12(\triangle {y})^2}
\left[-{\Phi}_{n-2}+16{\Phi}_{n-1}-30{\Phi}_n
\right.
\nonumber\\
&&\left.+16{\Phi}_{n+1}-{\Phi}_{n+2}\right]
-{\lambda}{\Phi}_n({\Phi}^2_n-1)\ .
\label{5_}
\end{align}
The ODE's (\ref{5_}) are 
solved using a forth-order Runge-Kutta scheme, and 
so our numerical algorithm is accurate to fourth 
order both in  time and in space, with error of 
 $O ((\triangle {y})^4)$ and $O ((\triangle {t})^4)$. 
For the boundaries, we set the left 
and right grid boundaries at ${y}_L=-40$ and 
${y}_R=+40$, 
and impose the condition as ${\Phi}({y}={y}_L,t)=
{\Phi}({y}={y}_R,t)=-1$ .
The grid number is $N=8000$
with the grid size of  
$\triangle {y}=1.0\times 10^{-2}$.
The initial position of a wall $y_0$ is set by 
${y}_0=|{y}_L+({y}_R-{y}_L+1)/3|=13 $, equivalently, 
one-third of the numerical range. 
For time steps, we set
$\triangle {t}=0.7\times \triangle {y}$. 

As for the particle production process, 
we have to solve 
the second-order ordinary differential equations (\ref{3_}) 
for each wave number ${k}$.
By using the fourth-order 
Runge-Kutta scheme, 
we solve them for the wave number $k$ of $0<k<100$ with the width 
$\triangle {k}=1.0\times 10^{-3}$. 
We estimate $|\beta_k|^2$ in the equation (\ref{6_}).
Defining the functions like these 
\begin{eqnarray}
W_1&\equiv&
\psi_k\dot{\psi}^{*}_k
-\psi^{*}_k\dot{\psi}_k\ ,\\
W_2&\equiv&
\psi_k\dot{\psi}^{*}_k
+\psi^{*}_k\dot{\psi}_k\ ,\\
W_3&\equiv&
\psi_k\psi_k^{*}\ ,
\end{eqnarray}
 we use the formula 
\begin{equation}
|\beta_k|^2=
\frac{(W_3-W_1/2i\omega)^2+(W_2/2\omega)^2}{4W_3}\ ,
\end{equation}
to evaluate $|\beta_k|$.
\section{\label{appen_b} Numerical examples of several bounces  at the 
collision}
\begin{figure}[ht]
 \centering
 \includegraphics[width=8cm]{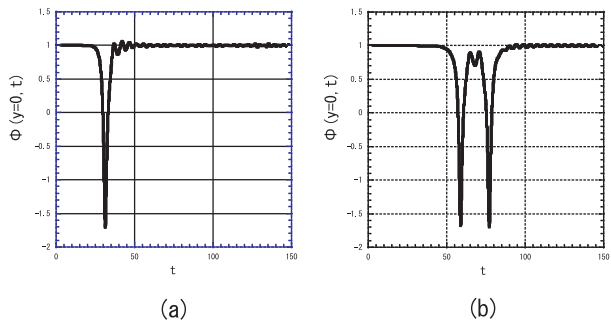}
\caption[fig1]{
Time evolution of  $\Phi$ field at $y=0$ are depicted for (a) 
$\upsilon=0.4 \ {\lambda}=1.0$
and (b) $\upsilon=0.2,\ {\lambda}=1.0$. 
The bounce occurs once for a large velocity (a), while 
Two bounces are
found for the slower velocity.  } \label{42}
\end{figure}

We depict some numerical examples which show several bounces at the 
collision. First we show two typical examples in Fig. \ref{42}.
\begin{figure}[ht]
 \centering
 \includegraphics[width=8cm]{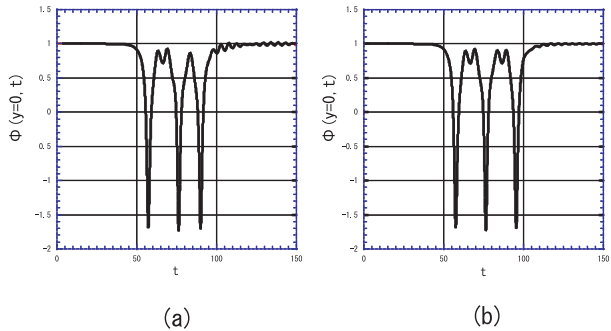}
\caption[fig1]{
Time evolution of  $\Phi$ field at $y=0$ are depicted for (a) 
$\upsilon=0.2062 \ {\lambda}=1.0$
and (b) $\upsilon=0.2049,\ {\lambda}=1.0$. 
Three bounces are found. 
} \label{21}
\end{figure}
\begin{figure}[hb]
 \centering
 \includegraphics[width=8cm]{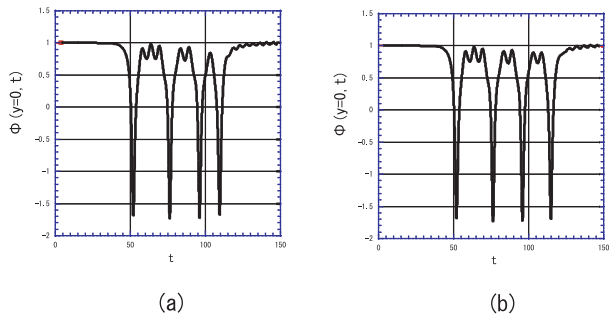}
\caption[fig1]{
Time evolution of  $\Phi$ field at $y=0$ are depicted for (a) 
$\upsilon=0.229283 \ {\lambda}=1.0$
and (b) $\upsilon=0.2292928,\ {\lambda}=1.0$. 
Four bounces are found. 
} \label{24}
\end{figure}
The figures show 
the behaviors of the scalar field $\Phi$ at $y=0$
with respect to $t$.
Initially, when two domain walls locate at large distance,
the value of the scalar field at $y=0$ is $1$.
Then two walls approach and collide.
At this point the value of $|\Phi-1|$ increases.
After the collision, it again decreases to an initial value.
We find some small oscillation around a domain wall structure which  is
excited by the collision.

From Fig. \ref{42}, we find there is one bounce for $\upsilon=0.4$, while 
a bounce occurs twice  for $\upsilon =0.2$.
In fact, the results are very much sensitive to the incident velocities 
as shown in
\cite{fractal}.

Here we present several examples to show how the behaviors of the scalar 
field  depend on $\upsilon$, which confirm the previous works.

In Figs.
\ref{21} (a) and (b), we show the case  for 
$\upsilon=0.2062$ and $\upsilon=0.2049$, respectively. 
 We find three bounces at the collision. 
Four-bounce solutions are depicted  in Figs. \ref{24} (a)
($\upsilon=0.229283$) and  (b) ($\upsilon=0.2292928$). 
These calculation shows that the detail collisional process is very sensitive to the
incident velocity.
\section{\label{perturb} perturbations of a domain wall}

We show that the  oscillation we find in Sec \ref{reheat2}, 
is a proper oscillation around a static stable domain wall.
To show it, we perturb the static domain wall solution (\ref{2_})
as
\begin{equation}
\Phi=\Phi_K(y)+\delta\Phi(t,y)\ .
\label{12_}
\end{equation}
In this appendix, we use the dimensionless variables rescaled by
$\eta$. Substituting this into Eq. (\ref{8}) and linearizing it, 
we obtain  
\begin{equation}
\delta\ddot{\Phi}-\delta\Phi''+\lambda(3\Phi_K^2-1)
\delta\Phi=0\ .
\label{8_}
\end{equation} 
Setting 
\begin{equation}
\delta\Phi=e^{-i\sigma t}F(y)\ ,\label{7_}
\end{equation}
and introducing new variable $\bar{y}=\tanh (y/d)$, 
we rewrite  
Eq.  (\ref{8_}) as 
\begin{equation}
(1-\bar{y}^2){d^2 F\over d\bar{y}^2}-2\bar{y}{d F\over d\bar{y}}
+2\left[3-\frac{2-\sigma^2/\lambda}{1-\bar{y}^2}
\right]F=0\ .
\end{equation}
The solution is given by the associated Legendre function.
Imposing the boundary condition ($F \to 0$ as $\bar{y}\to \infty$), 
we have two regular
solutions;
$F(\bar{y})=P_2^{\, 2}(\bar{y})$ with $\sigma=0$ and $P_2^{\, 1} (\bar{y})$
with $\sigma=\sqrt{3/2}\lambda^{1/2}$.

\ For the former mode, it just corresponds to a boost of 
a kink solution in the $y$ direction, because we find
$F(y)\approx \Phi_K(y)-\Phi_K(y-dy)$. 
The latter one is the oscillation mode which we find in Sec \ref{numer}.
In fact, taking the average over 10 cycles in the oscillation after the 
collision in  Fig. \ref{11}, we obtain the mean angular frequency is
about $1.17$, which is very close to  
$\sigma =\sqrt{3/2}\sim 1.22$.
The ratio is about 0.96.
We also evaluate the angular frequency for other cases.
We find  $1.33=1.09 \sigma$ for  Fig. \ref{12}, 
$3.67=0.95 \sigma$ for Fig. \ref{11.5},
and $4.18=1.08\sigma$ for Fig. \ref{13}. 
From the figures, we also find that
the amplitude of oscillation gets larger as
the incident velocity is faster.
This is because the excitation energy of a wall
 at the collision will be large for a large velocity.

We then conclude that the oscillations after the collision of two domain walls
is the proper oscillations around a stable domain wall.


\end{document}